\documentclass[aps,pra,twocolumn,groupedaddress,floatfix]{revtex4-2}

\usepackage{amsmath}
\usepackage{bm}
\usepackage{graphics}
\usepackage{times}
\usepackage{graphicx}
\usepackage{xcolor}
\usepackage{siunitx}
\usepackage{array,multirow,graphicx}
\usepackage{multirow,hhline,graphicx,array}
\usepackage{rotating}
\usepackage{multirow}
\usepackage{booktabs}
\usepackage[markup=underlined]{changes}
\begin{document}

\title{High-order harmonic generation in argon driven by short laser pulses: \\
effects of post-pulse propagation and windowing}
\author{Aaron T. Bondy}
\affiliation{Department of Physics and Astronomy, Drake University, Des Moines, Iowa 50311, USA}
\author{Klaus Bartschat}
\affiliation{Department of Physics and Astronomy, Drake University, Des Moines, Iowa 50311, USA}

\email{aaron.bondy@drake.edu}

\begin{abstract}
We present \textit{ab initio} calculations using the $R$-matrix with time dependence (\textsc{RMT}) method for high-order harmonic generation (HHG) in argon in a short, intense pulse regime.
The calculations employ a $6$-cycle $\sin^2$ pulse at $850$ nm with peak intensity $2.3\times 10^{14}$ W/cm$^2$ and, for comparison with the experiment by Guo \textit{et al.} [J.~Phys.~B: At.\ Mol.\ Opt. Phys.~{\bf 51}, 034006 (2018)], a Gaussian pulse with the same frequency and peak intensity.
Both pulse shapes yield the expected harmonic structure in the region above the ionization threshold (approximately $15.82$~eV in $LS$-coupling).
The spectra exhibit strong carrier-envelope-phase (CEP) sensitivity.
The energy region leading up to the ionization threshold contains spectral features arising from residual coherent dipole oscillations (free-induction decay) that strongly depend on spectral windowing and the post-pulse propagation time. 
We show that the HHG spectrum, particularly below the ionization threshold, is a defined quantity that depends on analysis choices rather than being a uniquely determined observable. 
Comparison between theoretical predictions and experimental observations in this energy regime, therefore, requires explicit specification of these parameters.
\end{abstract}

\maketitle

\section{Introduction}

High-order harmonic generation (HHG) arises when atoms are subjected to intense laser fields and is routinely used to generate coherent extreme-ultraviolet radiation and atto\-second pulses~\cite{Corkum1993,Lewenstein1994,Paul2001,KrauszIvanov2009}.
Corkum's three-step model~\cite{Corkum1993} describes HHG at a qualitative level in a semi-classical framework: an electron tunnel-ionizes, propagates in the continuum, and recombines with the parent atom, resulting in the emission of high-energy photons.
A more rigorous description is provided by the quantum-mechanical strong-field formulation of Lewenstein \textit{et al.}~\cite{Lewenstein1994}, which underlies much of modern atto\-second science~\cite{Paul2001,KrauszIvanov2009}.
The high-energy structure of HHG spectra is routinely characterized by the cutoff law, $E_{\mathrm{cut}} \approx I_p + 3.17\, U_p$, where $I_p$ and $U_p$ are the ionization and ponderomotive potentials, respectively, that relate the atomic and laser parameters to the cutoff energy, $E_{\mathrm{cut}}$.
In multi\-electron systems, however, this picture is incomplete.
Bound-bound coherence and post-pulse dynamics contribute significantly to the HHG spectrum, especially in the near-threshold region.

The quantitative characterization of HHG spectra in multi\-electron atoms, such as argon, requires a fully quantum-mechanical approach.
Although such treatment is possible and the predictions are observable, it is largely absent in the literature.
The main objective of almost all HHG calculations is to provide a \emph{qualitative} analysis of the high-energy plateau near the cutoff\,---\,the region responsible for the experimentally desired XUV pulse generation.

$R$-matrix with time-dependence (\textsc{RMT}~\cite{Brown2020}) is an \textit{ab initio} method to solve the time-dependent Schr{\"o}dinger equation (TDSE) for multi\-electron atoms and molecules that accounts for electron correlation and exchange within the region where these effects are expected to be significant.
\textsc{RMT} has been extensively and successfully applied to many strong-field processes including \hbox{HHG~\cite{clarke2018extreme,hassouneh2018cooper,Finger2022,Bondy2024high}}.
Comparisons between \textsc{RMT} and simpler approaches such as single-active-electron (SAE) models \hbox{\cite{Pauly2020,Finger2022,Bondy2024high}} that employ effective potentials \hbox{(e.g.,~\cite{Kulander1991,Tong2005})} generally reveal qualitative agreement for the harmonic structure, including the positions of the harmonics.
SAE approaches using pseudo\-potentials have been utilized to accurately reproduce essential physical features, such as the Cooper minimum in argon~\cite{Mauritsson2005}. 
These models often provide excellent physical intuition, without the need for a multi\-electron treatment, and can be very useful~\cite{Schafer2009}.
However, \emph{quantitative} agreement for measurable quantities such as the total emitted radiation (spectral fluence), either for the full spectrum or for a given harmonic, can differ significantly depending on the model and computational scheme~\cite{Finger2022}.

Within the few-pulse regime, additional effects must be considered.
A primary consideration is a marked increase in sensitivity to the carrier-envelope phase (CEP) of the driving laser. 
In short pulses, the CEP strongly influences electron trajectories and their interference, which determine the HHG spectrum~\cite{Guo2018,Pauly2020}. 
Additionally, the relative importance of short and long trajectories, and alterations in ionization timing can affect the harmonic yield and phase~\cite{Guo2018}.
CEP effects are important for comparison with experiment and therefore accounted for in the present work, although they are not the main focus.

A more fundamental concern involves the definition of the HHG spectrum itself\,---\,particularly its absolute magnitude~\cite{Finger2022,Bondy2024high}.
The HHG spectrum is related to the Fourier transform of the dipole moment induced by the laser-atom interaction.
Two parameters that influence this spectrum are the total propagation time\,---\,including field-free post-pulse propagation\,---\,and the potential application of windowing to the dipole moment or its time derivatives prior to taking the Fourier transform.
Experimentally, the total propagation time corresponds to the measurement window or the effective coherence time, whichever is shorter.
Previous studies have shown that post-pulse dynamics, in particular coherent bound--bound interactions between the ground and excited states, strongly influence the HHG spectrum, especially in the near-threshold region below the ionization limit~\cite{Camp2015,Beaulieu2016,Yun2018,peng2025resolving}. 
In particular, free-induction decay (FID) describes radiation emitted by coherent bound-state dynamics after the pulse has ended.

In TDSE simulations, these post-pulse contributions occur as a result of a coherent superposition of field-free eigenstates, which remain populated after the pulse and give rise to dipole oscillations at the transition frequencies.
Such calculations typically do not model intrinsic decoherence or spontaneous emission, especially with the latter occurring on a much larger timescale.
Consequently, the FID-like radiation does not decay after the pulse; rather, its spectral contribution increases in magnitude and sharpens with increasing propagation \hbox{time~\cite{Yun2018,peng2025resolving}}.
One of the purposes of windowing is to mitigate these effects~\cite{Bracewell2000,BlackmanTukey1959}.

As mentioned already, most HHG studies are focused on \emph{qualitative} features, including characterizing the high-energy portion of the plateau or the cutoff scaling, generally in service of attosecond pulse generation. 
As such, issues associated with FID-like emission are often not of primary importance.
However, when a \emph{quantitative} comparison with experimental spectra is desired, including a comparison of \emph{absolute} yields, these effects\,---\,and therefore knowledge of windowing and post-pulse propagation time\,---\,become essential.
Recent studies have shown that different theoretical models and choices of the dipole gauge can also have an impact on the resulting spectra~\cite{Finger2022,Bondy2024high}.

The HHG spectrum is extracted from a finite-time dipole signal, representing the radiation emitted over that chosen window; it is not a uniquely defined, time-independent observable.
Thus, quantities such as the spectral density or total emitted energy depend explicitly on how and whether the induced dipole is windowed and also on the simulation duration, which determines how much FID-like radiation contributes to the resulting spectra~\cite{Bracewell2000,Harris1978}.

In the present study, we use the \textsc{RMT} method to investigate HHG in the strong-field, few-cycle regime in argon, using a parameter set that is closely comparable to the joint experimental-theoretical work of Guo \textit{et al}.~\cite{Guo2018}.
We benchmark our above-threshold harmonic features against their results, though some differences remain since those authors imposed selection mechanisms to choose only short trajectories, whereas we make no such restriction in the present study.

The primary objective of our work, beyond presenting absolute HHG spectra for this set of parameters in argon, is to quantify how the near-threshold Rydberg region of the spectrum depends on post-pulse propagation duration and spectral windowing.
We show that windowing affects both the detailed spectral structure and the total integral of the spectral distribution\,---\,especially in the region leading up to the ionization threshold that represents the radiation emitted from the persistent dipole oscillations.
Although CEP effects are independent of these two concerns, they are still analyzed in the present study.
We aim to provide insight into the interpretation of HHG spectra in the short-pulse regime and to establish a framework for consistent and quantitative comparison with experiment, particularly in multi\-electron targets within the near-threshold energy region~\cite{Higuet2011,Gibson2004}.

\section{Methods}

We use the $R$-matrix with time-dependence (\textsc{RMT}) method to solve the time-dependent Schrödinger equation (TDSE). \textsc{RMT} is an \textit{ab initio} approach widely used to treat the case of multi\-electron targets interacting with intense laser fields~\cite{Brown2020}. 
In atomic units, which are used throughout this paper unless stated otherwise, the TDSE is written as
\begin{equation}
i\frac{\partial}{\partial t}\Psi(\mathbf{r},t) = \left[ H_0 + V(\mathbf{r},t) \right]\Psi(\mathbf{r},t).
\end{equation}
$H_0$ is the field-free Hamiltonian and $V(\mathbf{r},t)$ describes the time-dependent laser-atom interaction, treated here in the semiclassical dipole approximation, as is standard in this regime~\cite{Guo2018,Brown2020}. 
\textsc{RMT} uses the $R$-matrix paradigm, separating the configuration space into an inner region, where all-electron effects including exchange and correlation are treated fully, and an outer region containing at most one electron.  The latter is far enough away from the residual ion that exchange effects with the other electrons can be ignored.

\textsc{RMT} deploys a different numerical scheme in the two regions. 
Within the inner region, the wave function is expanded in a close-coupling basis of residual-ion target states coupled to orbitals of the additional electron (both bound and continuum), with the radial part represented by a $B$-spline basis.
In the outer region, the wave function is constructed as a close-coupling expansion between channel functions (the residual-ion states coupled to the angular degrees of freedom of the ejected electron), and the reduced radial wave functions for each channel, which are represented on a finite-difference radial grid.
Contrary to other $R$-matrix approaches, the wave function itself, rather than the $R$-matrix, is matched explicitly at the inner-outer region boundary in \textsc{RMT}.

In the present calculations, we employ an $R$-matrix radius of 20, a grid spacing in the outer region of 0.08, and a time-step of 0.01\,---\,parameters known to work well in previous \textsc{RMT} calculations~\cite{Pauly2020,Finger2022,Bondy2024high}, but checked again for confirmation here. 
The appropriate size of the outer region (to prevent spurious reflections) can be obtained from the classical excursion length $R_{\mathrm{max}} = \frac{2 E_0}{\omega^2}(N_{\mathrm{cycles}}\pi + 1)$, where $E_0$ and $\omega$ are the peak field amplitude and frequency of the laser, respectively.
This ensures that the wave function for the active electron remains confined within the entire (inner plus outer) box for the entire simulation. We do not utilize any absorbers in the present work.
We found that $R_{\mathrm{max}}$ ($\approx$ 1120 a.u.) is sufficient for pulse-length simulations \hbox{($T\approx 703~\rm{a.u.} =17.0$~fs)}. 
We also carried out simulations that lasted beyond the length of the pulse, in order to study continued post-pulse effects.
For the longest simulation time \hbox{($\approx 4218~\rm{a.u.} =$ 102~fs)}, where the simulation continues for five additional pulse durations beyond the pulse itself, $R_{\mathrm{max}} \approx 5160$ was found to be adequate. 
In the inner-region, the continuum functions were constructed by employing 50 $B$-splines with a spline order of~13.

\subsection{Atomic structure and target description}

The present calculations employ the Ar$^+$ target structure of Burke and Taylor~\cite{burke1975r}.
The $(N\!+\!1)$-electron close-coupling expansion contains the $3s^2 3p^5 \epsilon \ell$ and $3s 3p^6 \epsilon \ell$ channels, including doubly-excited configurations of the residual ion that accurately chracterize the $3s3p^6 np$ window resonances~\cite{Finger2022}. 
Total angular momenta up to $L_{\max}=50$ were coupled to ensure numerically converged results.

\subsection{Laser field and computational parameters}
\label{sec:IIB}
The argon atom is subjected to a short infrared laser pulse with a central wavelength $\lambda = 850$ nm and peak intensity $I_0 = 2.3\times 10^{14}$ W/cm$^2$, using primarily a 6-cycle $\sin^2$ envelope, and for one simulation a Gaussian envelope with a comparable 6.2~fs full width at half maximum (FWHM) intensity.
The Gaussian pulse allows close comparison with Ref.~\cite{Guo2018}, as shown in Fig.~1.

In the present work, the electric field is linearly polarized along the $z$ axis and given by
\begin{equation}
\mathbf{E}(t) = f(t)\,E_0 \cos(\omega t + \phi)\,\hat{z}.
\end{equation}
Here $E_0$ is the peak field amplitude (related to the peak intensity via $I_0 = \frac{1}{2}\,\epsilon_0 c\,E_0^2$), $\omega$ is the laser frequency, $\phi$ is the carrier-envelope phase, and $f(t)$ is the pulse envelope. 
The $\sin^2$ envelope is
\begin{equation}
f(t)=\sin^2\!\left(\frac{\omega t}{2N}\right), \quad 0 \le t \le \frac{2\pi N}{\omega},
\end{equation}
where $N$ is the number of cycles. 
As mentioned above, the ``equivalent'' Gaussian has an intensity FWHM of 6.2~fs.
The importance of CEP effects, including CEP-averaging, is assessed by performing calculations for several values of $\phi$ and averaging over them when appropriate~\cite{Guo2018,Pauly2020}.

\subsection{Observables and spectral extraction}

The important observable in the HHG process is the time-dependent dipole moment, 
\begin{equation}
d(t) = \langle \Psi(t) | z | \Psi(t) \rangle,
\end{equation}
where the laser polarization axis lies along $\hat{\mathbf{z}}$.

The HHG spectrum is then obtained via the Fourier transform of the dipole acceleration,
\begin{equation}
    \label{eq:eq5}
\tilde{a}(\omega) = \int_{t_i}^{t_f} a(t)\, e^{i\omega t}\, dt,
\end{equation}
from which the spectral density is defined~\cite{Joachain2012,telnov2013exterior} as
\begin{equation}
    \label{eq:eq6}
S(\omega) = \frac{2}{3\pi c^3} \left| \tilde{a}(\omega) \right|^2.
\end{equation}
This form of $S(\omega)$ is consistent with previous \textsc{RMT} implementations~\cite{Finger2022,Bondy2024high}. $S(\omega)\,d\omega$ is the energy emitted into the frequency interval $(\omega, \omega + d\omega)$, and the integral $\int S(\omega)\,d\omega$ is the spectral fluence~\cite{Jackson1999}.

The limits of integration $(t_i,t_f)$ in the Fourier transform in Eq.~\eqref{eq:eq5} define the \textit{finite} time window for the full simulation\,---\,the laser pulse and a variable post-pulse propagation duration. 
This interval is consequential, as it determines the degree to which the long-lived dipole oscillations influence the resulting HHG spectra in the near-threshold region below the ionization limit. 
This process is known as free-induction decay (FID)~\cite{Beaulieu2016} and is distinct from spontaneous decay.
Specifically, extending the length of the simulation increases the signal obtained from residual post-pulse oscillations. 
Therefore, the HHG spectrum\,---\,in particular the features corresponding to ground-excited dipole oscillations\,---\,is sensitive to the simulation duration.

For definiteness, the present study employs $LS$-coupling, so the residual-ion $3s^2 3p^5\,{}^2P^o$ ground state lacks fine-structure resolution and is regarded as a single term.
The ionization limit is taken as the term centroid, approximately 15.82~eV, not the (usually quoted) lowest fine-structure threshold at 15.76~eV~\cite{kramida2024nist}. 
In the present manuscript, references to the ionization threshold should be understood to refer to the term centroid.

While the different forms (acceleration, velocity, length) of the dipole operator give consistent spectra when the calculation is converged, their definitions differ by factors of $\omega^2$ and $\omega^4$ and the results may exhibit deviations for finite-time propagation. 
The acceleration form is generally preferred, especially for short pulses~\cite{Bandrauk2009,Han2010}, and is used in the present work.

Windowing mitigates spectral leakage and spurious oscillations that arise from discontinuities at the boundaries of the interval used in the Fourier transform~\cite{Camp2015,Harris1978}. The technique is particularly useful in HHG calculations for characterizing high-energy features in the spectrum near and just beyond the cutoff. 
However, as a by-product, it alters the spectral fluence over the full spectrum\,---\,a quantity that is, in principle, experimentally measurable~\cite{Finger2022,Bondy2024high}. 
This modification is especially pronounced in the below-threshold Rydberg region, as will be shown in the following section. 
In this regime, spectral features arising from persistent post-pulse dipole oscillations are highly sensitive to both the choice of windowing and propagation time~\cite{Camp2015,Yun2018,peng2025resolving}.

To quantify the latter aspect, in certain simulations, the wave function is propagated beyond the duration of the pulse in order to examine how the calculated spectral fluences corresponding to particular ground-excited dipole oscillations depend on the total propagation time.
Immediately after the pulse, the wave packet of the ejected electron can still recombine, thereby leading to emission and even recapture into excited states.
Over longer time periods, the coherent dipole oscillations between the ground and accessible excited states lead to radiation at the transition energies of these states.
Our simulations do not include any decay or decoherence mechanisms. Therefore, these contributions are persistent and increase over time\,---\,as they would be expected to do experimentally until decoherence suppresses them.

\section{Results and Discussion}
\label{sec:results}

\subsection{Benchmarking against Guo \textit{et al.}}
\label{subsec:benchmark}
 
Our first objective is to compare our HHG spectra to the Guo \textit{et al.}~\cite{Guo2018} experiment, which employed very similar laser parameters.
The present simulations, beginning with a comparison to Ref.~\cite{Guo2018}, are carried out using the parameters detailed in Sec.~\ref{sec:IIB}. 
These parameters constitute a rather unusual HHG regime, as they induce about a $10\%$ ground-state depletion in argon. 

\begin{figure}[t]
\centering
\includegraphics[width=0.48\textwidth]{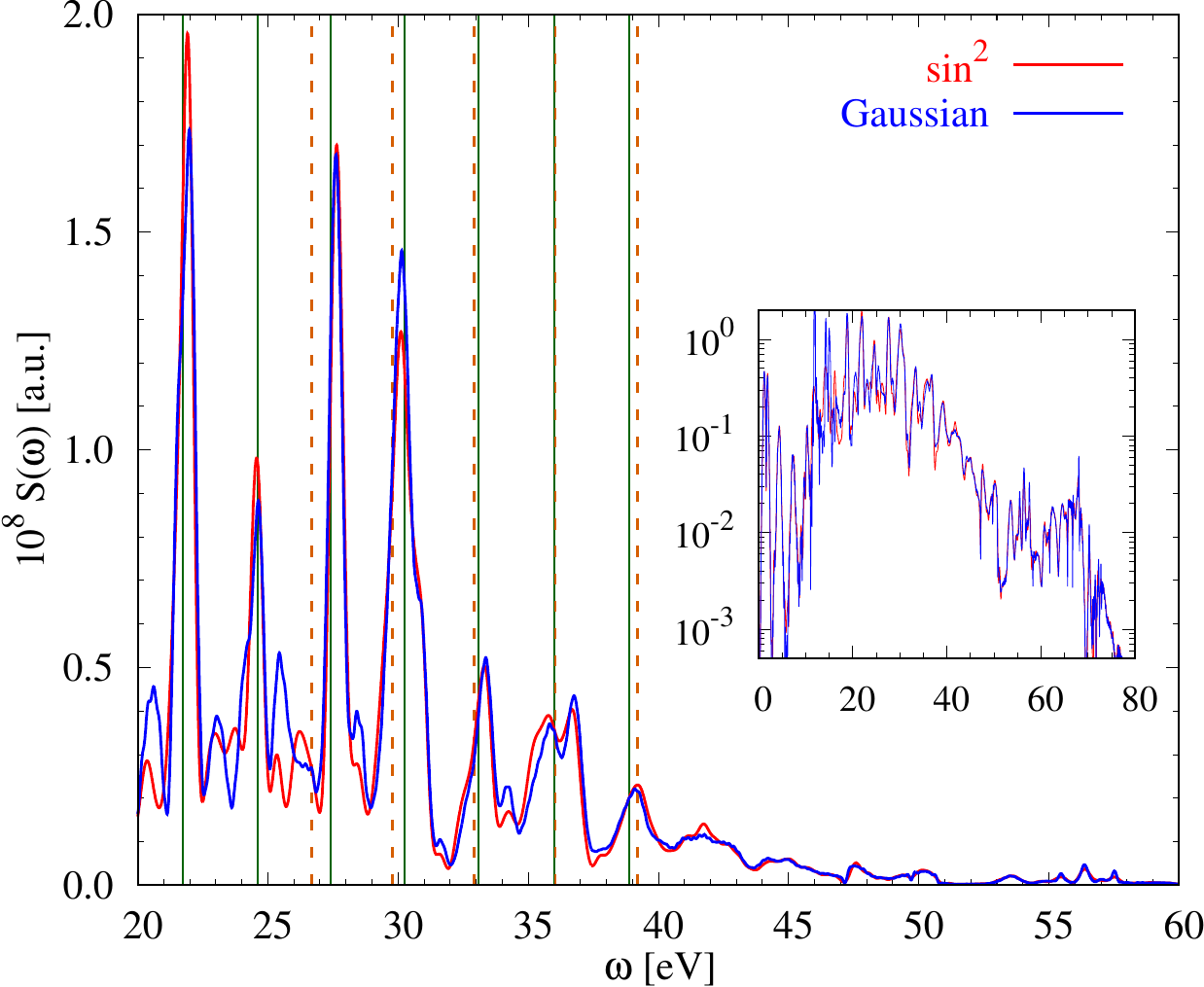}
\caption{
CEP-averaged HHG spectra calculated with both the $\sin^2$
and Gaussian pulses described in Sec.~\ref{sec:IIB}. The $20$--$60$~eV range is shown on a linear scale, with comparisons to the peak harmonic positions in the theoretical (vertical solid green line) and experimental (vertical dashed orange line) work of Guo \textit{et al.}~\cite{Guo2018}. The inset shows the full HHG spectrum from 0 to 80~eV on a logarithmic scale.
}
\label{fig:1}
\end{figure}

In Fig.~\ref{fig:1}, CEP-averaged HHG spectra for both pulse types are shown between $20$ and $60$~eV and compared to the results of Guo \textit{et al.}~\cite{Guo2018}. 
To begin with, we note (cf.\ the inset of Fig.~\ref{fig:1}) that the two pulse shapes used in the present calculation yield very similar results. 
It can be seen that the positions of the dominant peaks for both pulse types are consistent with the harmonic positions calculated and observed in Ref.~\cite{Guo2018}, indicated by the solid green (theory) and dashed orange (experimental) vertical lines.
Overall, the present calculations show good agreement with the experimental results of Guo \textit{et al.}~\cite{Guo2018}, with particularly close agreement for features near 30 and 39~eV. For peaks near 26.7 and 33~eV, the theoretical results of Ref.~\cite{Guo2018} agree more closely with their experimental data than the current ones, although the overall level of agreement across the spectrum is comparable. 

Nevertheless, the comparison with Ref.~\cite{Guo2018} is somewhat limited. 
Firstly, Guo \textit{et al.}~\cite{Guo2018} used a setup with a stable but unknown CEP. 
Furthermore, they employed a high-pressure argon gas jet with conditions that favor short HHG trajectories.  Consequently, they selected these trajectories by employing absorbers within their single-active-electron calculations. 
The present calculations employ no such trajectory-selection restrictions.  

Another caveat pertaining to a direct comparison with Ref.~\cite{Guo2018} concerns the pulse itself. 
Although we have chosen a Gaussian pulse with identical parameters to those of their TDSE calculations, their experimental pulse, to the extent that it can be accurately spectrally decomposed, exhibits a residual chirp and some temporal asymmetry. 
HHG spectra in the few-cycle regime are known to be very sensitive to these parameters~\cite{Lewenstein1994,Guo2018,Camp2015,Yun2018}. 
All these differences inform our decision to resist a more detailed analysis and to simply regard the favorable comparison of the peak positions of the harmonics as a meaningful benchmark. 
From here on, all exhibited results were obtained from calculations with a $\sin^2$ pulse.

\subsection{CEP sensitivity}
\label{subsec:cep}

\begin{figure}[t]
\centering
\includegraphics[width=0.48\textwidth]{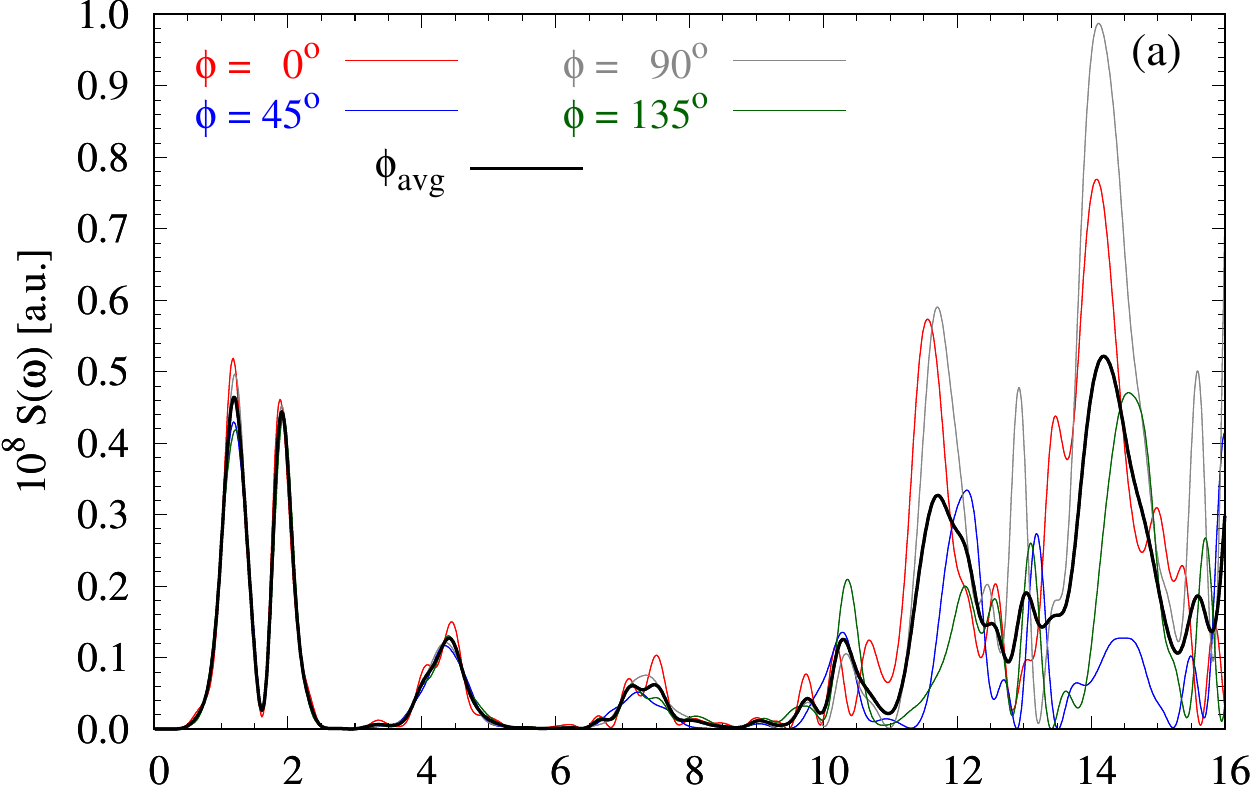}

\vspace{0.1cm}

\includegraphics[width=0.48\textwidth]{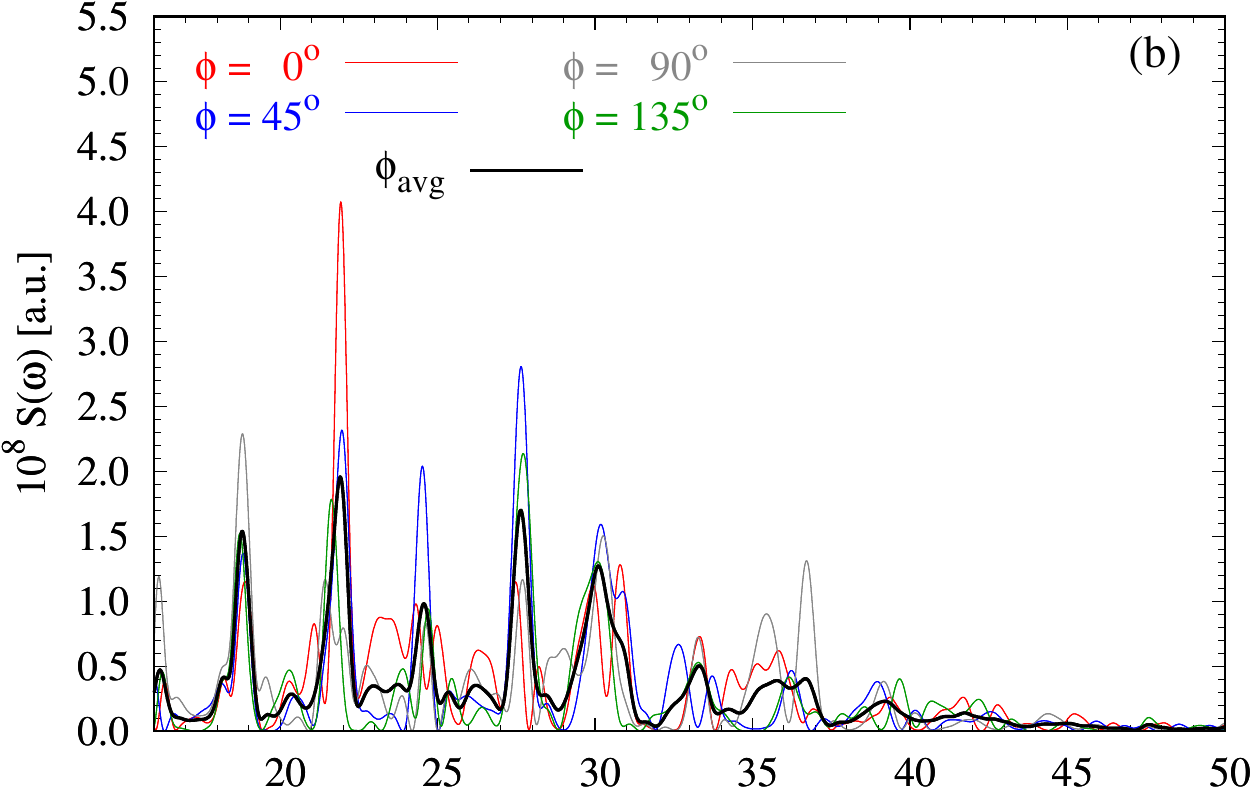}

\vspace{0.1cm}

\includegraphics[width=0.48\textwidth]{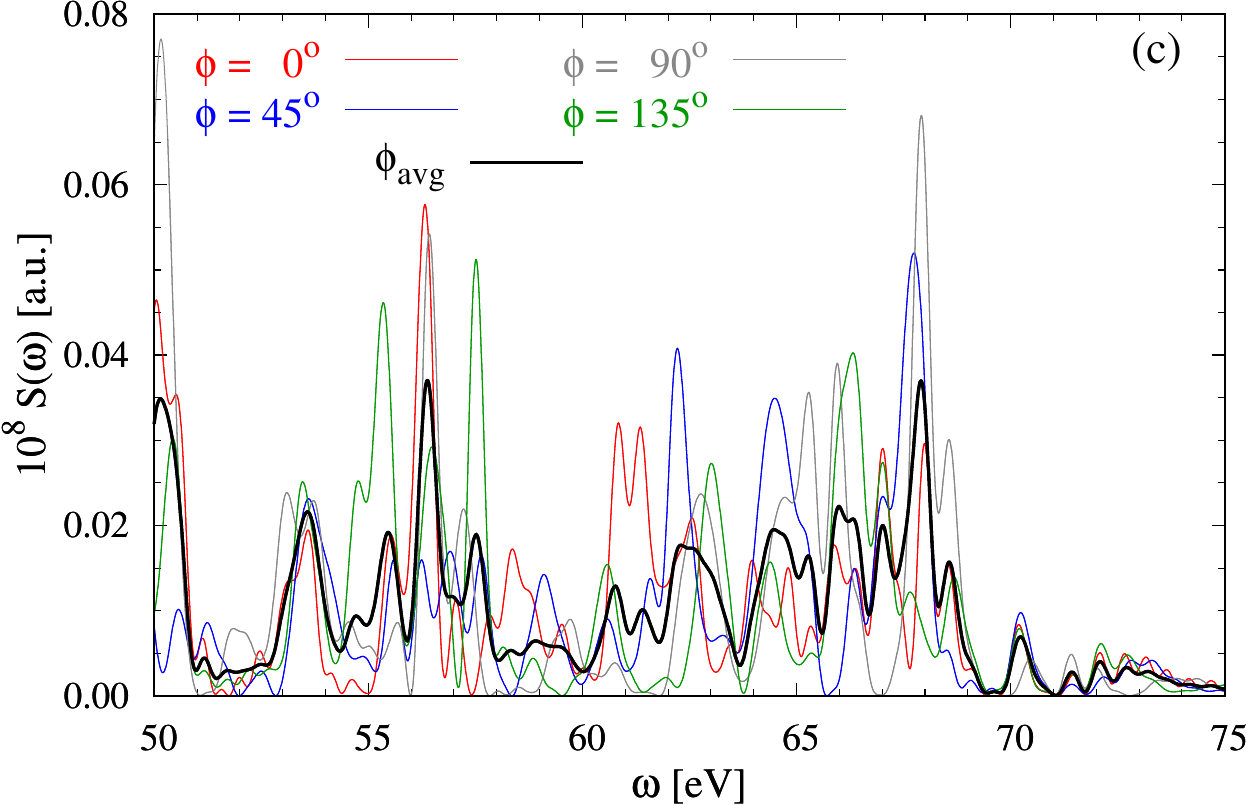}

\caption{
CEP-resolved ($\phi=$ $0^\circ$, $45^\circ$, $90^\circ$, and $135^\circ$) and CEP-averaged HHG spectra shown for (a) the below-threshold \hbox{($E<15.82$~eV)} region, (b) the above-threshold region between 16 and 50~eV, and (c) the high-energy tail between 50 and 75~eV.
}
\label{fig:2}
\end{figure}

Figure~\ref{fig:2} demonstrates the strong dependence of the HHG spectra on the CEP by showing simulations carried out for four characteristic CEPs, $\phi=0^\circ$, $45^\circ$, $90^\circ$, and $135^\circ$, along with a CEP-averaged spectrum. 
As mentioned above, Guo \textit{et al.}~\cite{Guo2018} employed a stabilized but unknown CEP, introducing variations via dispersion through a glass wedge.
Panel~(a) shows the region below the ionization threshold \hbox{($\approx$ 15.82~eV)}, panel~(b) displays the energy region above this limit up to 50~eV, and panel~(c) exhibits the high-energy tail between 50 and 75~eV. 
The spectra, except for the first few harmonics below 5~eV, are highly sensitive to the CEP, as is expected for such short pulses~\cite{Guo2018,Yun2018}.
A notable exception is in the above-threshold energy region between approximately 18 and 30~eV, shown in panel~(b), where the peaks of the harmonics are found to coincide for all of the CEPs. 
The experiment of Guo \textit{et al.}~\cite{Guo2018} similarly identified this spectral region where the harmonic response is only weakly dependent on the CEP of the driving laser, a feature they exploited in shaping the corresponding spectral phase in attosecond pulse trains. 

In contrast, the region between 10 and 16~eV exhibits a strong CEP-sensitivity, due to dynamics involving the near-threshold excited-state manifold. 
Although this particular simulation was only carried out for the duration of the pulse, features corresponding to dipole oscillations already appear at the bound-excited transition energies between 10 and 16~eV. 
The transition energies are Stark-shifted and broadened here, but a comparison can be made to Fig.~\ref{fig:6} below, which contains several pulse durations.
As discussed in Sec.~\ref{subsec:rydberg}, these features will grow substantially after the pulse ends, as the coherent dipole oscillations persist once the driving field is turned off.

As a sidenote, when we examine windowing in the following Sec.~\ref{subsec:windowing_def}, we use CEP-averaged spectra. However, when studying post-pulse effects using longer simulation times in Sec.~\ref{subsec:rydberg}, we choose a single CEP ($\phi=0$), since the calculations become time-consuming and computationally expensive. 
More importantly, both of these effects\,---\,post-pulse dipole oscillations and windowing\,---\,are CEP-independent and hence can be demonstrated without CEP-averaging.

\subsection{Windowing and spectral extraction}
\label{subsec:windowing_def}

In HHG studies, windows are applied to a finite-duration signal before taking its Fourier transform, with the purpose of suppressing spectral artifacts and mitigating the effects of post-pulse emission~\cite{Bracewell2000,Harris1978}.
Mathematically, windowing amounts to replacing $a(t)$ with $w(t)a(t)$, where $w(t)$ is defined over the chosen time interval $[0,T]$, which includes the pulse and any post-pulse propagation.

Although typical attosecond applications focus on the plateau and cutoff regions, where windowing can mimic the physically relevant emission time window defined by recollision, the choice of window function has non-trivial consequences for other parts of the spectrum.
It affects the spectral density and the (total and harmonic-specific) fluence, both of which are measurable quantities.
These effects are most pronounced in the near-threshold excited-state region, where long-lived dipole oscillations are suppressed by windowing.

\begin{figure}[b]
\centering
\includegraphics[width=0.48\textwidth]{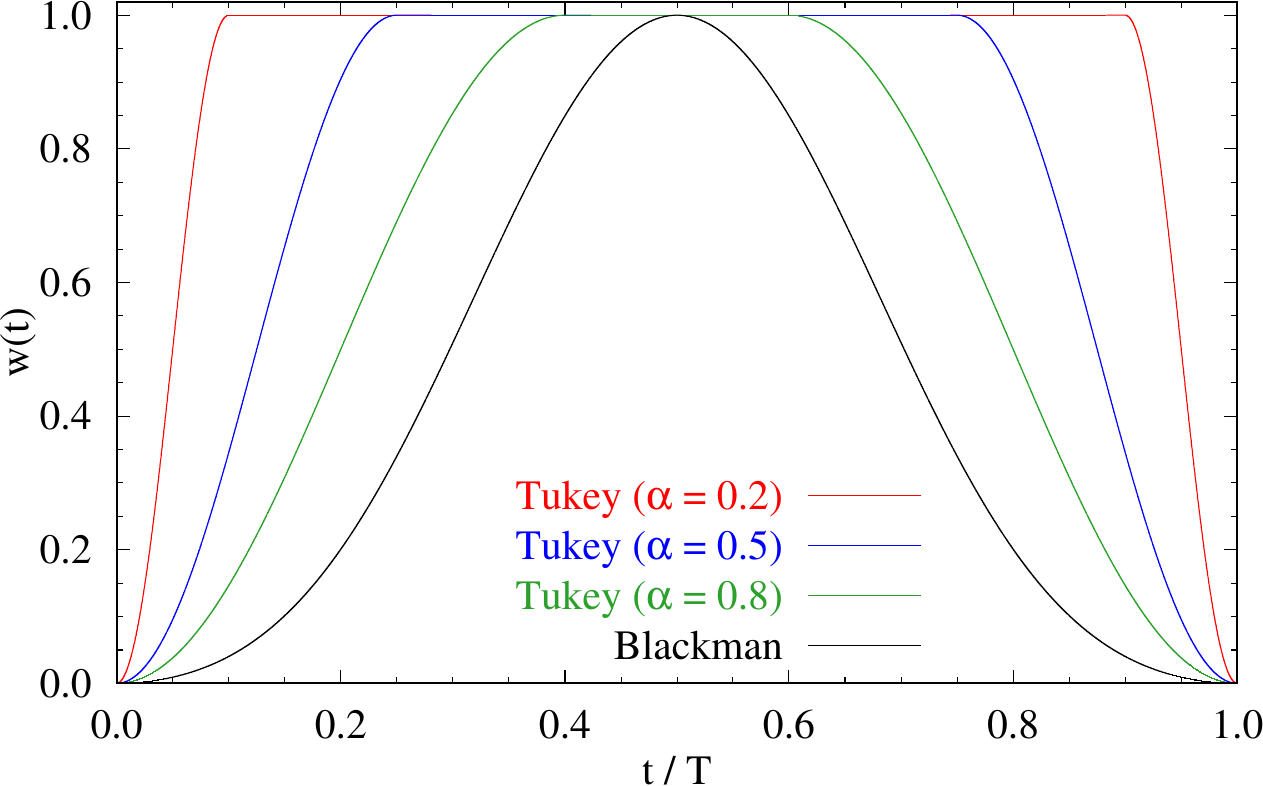}
\caption{
Illustration of the window functions used in this paper. The Blackman window, corresponding to Eq.~\eqref{eq:w-blackman}, and three Tukey windows ($\alpha=$ 0.2, 0.5, and 0.8), corresponding to Eq.~\eqref{eq:w-tukey}, are plotted as a fraction of the simulation time, $T$.
}
\label{fig:3}
\end{figure}

The first of two windows we employ is the Blackman window~\cite{BlackmanTukey1959}, defined as
\begin{equation}
\label{eq:w-blackman}
w_{\mathrm{B}}(t)
=
0.42
-0.5\cos\bigl(\tfrac{2\pi t}{T}\bigr)
+0.08\cos\bigl(\tfrac{4\pi t}{T}\bigr),
\quad 0 \le t \le T.
\end{equation}
As seen in Fig.~\ref{fig:3}, the Blackman window strongly suppresses the edges of the signal and does not have a flat central region, compared to other windows.
Extending beyond reducing numerical artifacts, this window also suppresses late-time emission.
In this way, it significantly limits the contributions of post-pulse dynamics.

The other window we employ is the Tukey window~\cite{Harris1978}\,---\,a tapered cosine function characterized by the parameter \hbox{$\alpha \in [0,1]$}, which allows a variable width of the central flat region and thus selective suppression of numerical artifacts and post-pulse effects.
It is defined by
\begin{equation}
    \label{eq:w-tukey}
w_{\mathrm{T}}(t)=
\begin{cases}
\frac{1}{2}\left[1+\cos\bigl(\pi(\tfrac{2t}{\alpha T}-1)\bigr)\right],
& {\scriptstyle 0\, \le \, t \, < \tfrac{\alpha T}{2}}, \\[3pt]
1,
& {\scriptstyle \tfrac{\alpha T}{2} \le \, t \, \le \, T\left(1-\tfrac{\alpha}{2}\right)}, \\[3pt]
\frac{1}{2}\left[1+\cos\bigl(\pi(\tfrac{2t}{\alpha T}-\tfrac{2}{\alpha}+1)\bigr)\right],
& {\scriptstyle T\left(1-\tfrac{\alpha}{2}\right) \, < \, t \, \le \, T}.
\end{cases}
\end{equation}
Figure~\ref{fig:3} shows three Tukey windows for $\alpha=0.2$, 0.5, and 0.8, respectively.

\begin{figure}[t]
\centering
\includegraphics[width=0.48\textwidth]{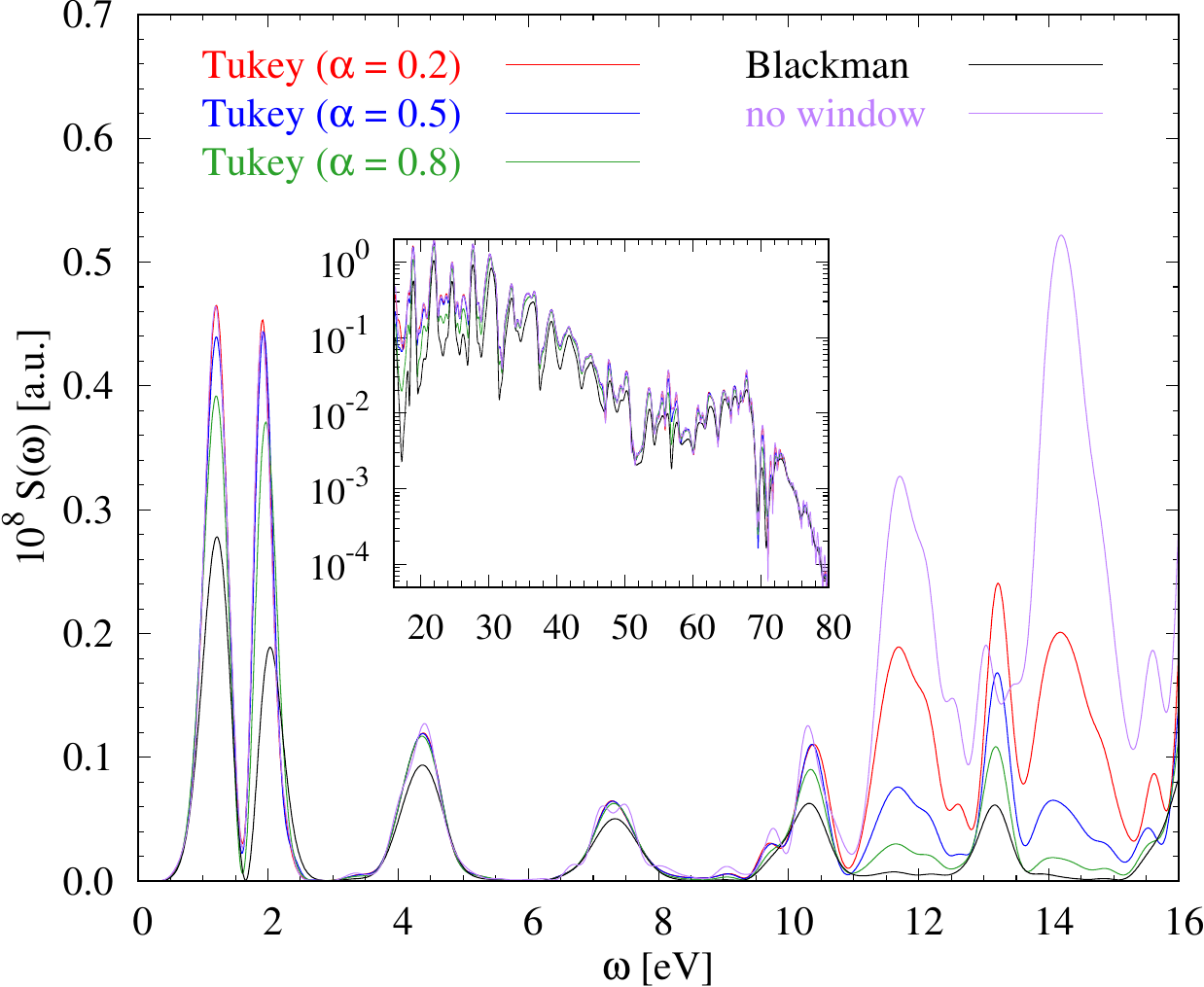}
\caption{
CEP-averaged HHG spectra obtained by applying the Blackman and three Tukey windows (with $\alpha=$ 0.2, 0.5, and 0.8), shown in Fig.~\ref{fig:3}, to the dipole acceleration, $a(t)$, in Eq.~\eqref{eq:eq5}, before taking the Fourier transform.
The ``no window'' case is included for comparison. 
 The below-threshold energy region is shown on a linear scale, while the inset displays the above-threshold region on a logarithmic scale. 
}
\label{fig:4}
\end{figure}

While the Blackman window has the most extreme effect on the spectrum, the Tukey windows impact the spectrum less as $\alpha$ decreases, since the unmodified central region becomes larger. In the limit $\alpha \to 0$, the Tukey window reduces to the rectangular (no-window) case.
The absence of a window (labelled ``no-window'' in Fig.~\ref{fig:4}) corresponds to abrupt truncation of the signal at the endpoints.
Like the windowed cases, this is also an approximation.
In spite of this, to quantify spectral fluence as accurately as possible, the signal should be left unwindowed, since windowing suppresses the very feature of interest\,---\,radiation from coherent dipole oscillations\,---\,that we are studying.

Figure~\ref{fig:4} exhibits the effect of windowing on the HHG spectra for all four windows as well as for the unwindowed case.
A suppression in the spectral density is observed in the windowed spectra, relative to the unwindowed case, at almost all frequencies\,---\,though especially in the near-threshold excited-state region between about 12 and 16~eV.
As expected from its functional form, the Blackman window causes the strongest suppression and is useful in many HHG applications for reducing spectral artifacts; however, for studying this region, the unwindowed case is clearly preferable.
The Tukey windows produce less suppression than the Blackman window, with $\alpha = 0.2$ yielding results closest to the unwindowed case.
At isolated frequencies, the Tukey ($\alpha = 0.2$) spectrum slightly exceeds the unwindowed result. This arises from a redistribution of spectral weight due to the modified finite-time interference pattern introduced by the window, rather than any physical enhancement of the emission.
Since the harmonic peaks are well-separated and sufficiently strong (i.e., above the background) in the below-threshold region, windowing provides only minimal noise-reduction benefit while instead distorting the true spectral density.
The unwindowed case is thus the most faithful representation of the physical spectrum in this energy region.

\subsection{Bound-excited post-pulse coherence}
\label{subsec:rydberg}

We next investigate the near-threshold Rydberg region, extending from the first excited state feature up to the $LS$-coupled ionization limit of approximately $15.82$~eV.
The spectral features in this region are identified as free-induction decay (FID) and accumulate during and especially after the pulse.
As mentioned above, for this analysis we use a single CEP ($\phi = 0$) to conduct simulations beyond the length of the pulse itself. Recall that this effect is independent of the CEP.

\begin{figure}[t]
\centering
\includegraphics[width=0.48\textwidth]{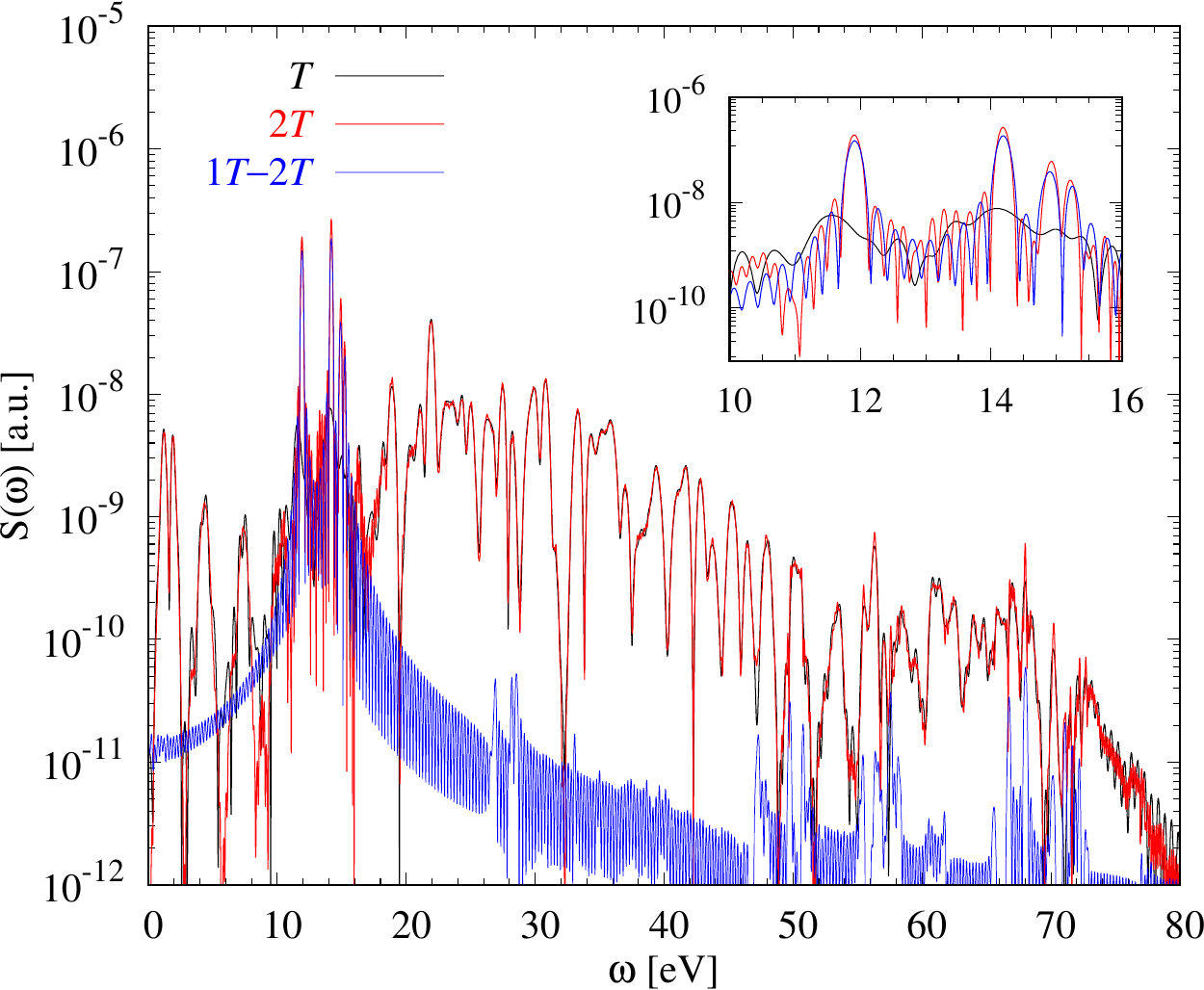}
\caption{
CEP-resolved HHG spectra ($\phi=0$) shown for simulation times equal to the pulse duration ($T$), twice that amount ($2T$), and also for the spectrum resulting exclusively from the post-pulse period, $T<t<2T$.
}
\label{fig:5}
\end{figure}

\begin{figure}[t]
\centering
\includegraphics[width=0.48\textwidth]{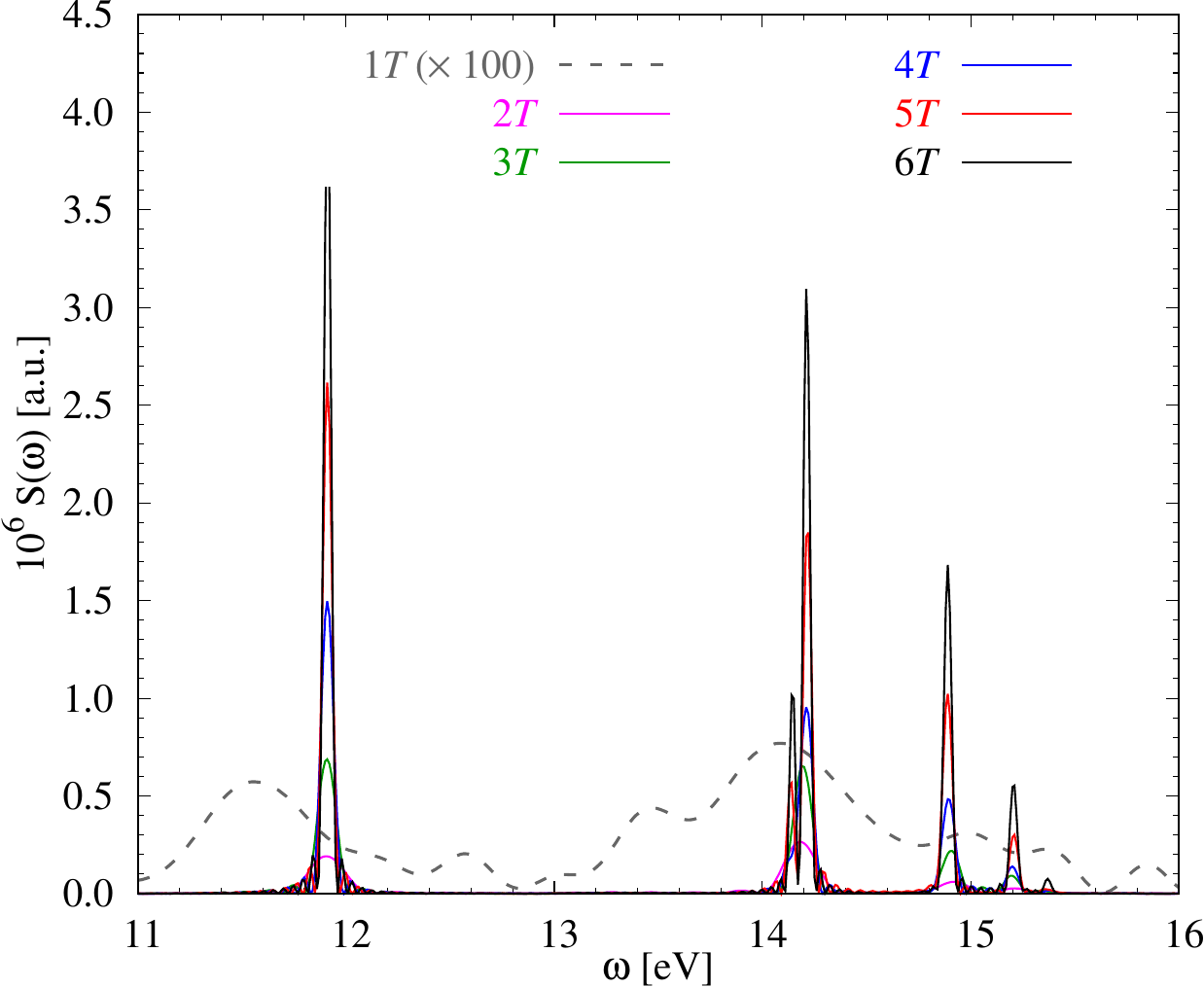}
\caption{
HHG spectra in the near-threshold region arising from ground-to-excited-state dipole coherences, shown for simulation times ranging from the pulse duration alone ($1T$, multiplied by 100 for visibility) to increasing post-pulse propagation times up to five times the pulse duration ($6T$). The excited states and the integrals under their spectral features are listed in Table~\ref{tab:int_scaled}.
}
\label{fig:6}
\end{figure}

Figure~\ref{fig:5} illustrates the effect of propagating the wave function one pulse duration beyond the original laser pulse by showing the HHG spectra obtained after (i) just the pulse ($T$), (ii) the pulse + an extra pulse length of propagation ($2T$), and (iii) the spectra from only the extra propagation period \hbox{($1T$--$2T$)}. 
Looking at the near-threshold region (approximately the region shown in the inset), it is clear that the post-pulse dynamics are the dominant contributions to these FID effects, which arise from coherent dipole oscillations between the accessible excited states (populated during the pulse) and the ground state.
That these features appear at the field-free frequencies, as shown more clearly in Fig.~\ref{fig:6} and Table~\ref{tab:int_scaled}, supports the interpretation that they are the result of post-pulse effects.
While this phenomenon already begins during the pulse itself, the inset shows that these features appear at slightly Stark-shifted frequencies and are broadened, as they compete with recollision-driven dynamics.
Furthermore, although several bound--bound dipole coherences are present in the spectrum, the substantial survival of the ground state ensures that ground-to-excited-state contributions of the form $\langle {}^1S^e | z | {}^1P^o \rangle$ dominate the post-pulse HHG spectrum.
These correspond to transitions between the ground state and the $3p^5 n\ell\,{}^1P_1^o$ states.

At certain energies above 25~eV, the post-pulse-only \hbox{($1T$--$2T$)} spectrum depicted in Fig.~\ref{fig:5} exhibits distinct features above the background, though still well below the full (pulse + propagation) HHG spectra ($2T$).
These features arise from recombination in the immediate aftermath of the pulse and persist even in the presence of windowing.  

Figure~\ref{fig:6} displays the HHG spectra in the $11.8$--$15.82$~eV near-threshold region for various simulation times, ranging from the pulse duration alone ($1T$) to the pulse followed by  additional propagation times of up to five times the pulse duration ($6T$).
The sharp peaks observed in the spectra correspond to the field-free transition energies between the ground and the dipole-accessible excited states.
The dipole coherences cause radiation at these frequencies, which accumulates after the pulse.
We see that these effects are significantly broader (and much smaller in magnitude) when we consider solely the spectrum collected over the pulse duration, though significant dipole coherence contributions remain on this time scale, albeit shifted.

\begin{table}[t]
\small
\setlength{\tabcolsep}{3pt}
\caption{
Spectral fluences, $\int S(\omega)\,d\omega$, multiplied by $10^{8}$, over the energy ranges in Fig.~\ref{fig:6} corresponding to dipole coherences between the ground state, $3p^6\,{}^1S^e$, and the $3p^5\,4s$, $3p^5\,3d/5s$, and $3p^5\,4d$ states (all ${}^1P^\circ$ symmetry) of neutral argon.
$T_p$ ($T$) denotes the pulse (simulation) duration.
Blank entries correspond to spectral features that are not yet well-resolved.
The columns labeled $\Delta$ give the finite differences between successive fluences. 
}
\label{tab:int_scaled}
\begin{ruledtabular}
\begin{tabular}{c S[table-format=2.3] S[table-format=2.3] S[table-format=2.3] S[table-format=2.3] S[table-format=2.3] S[table-format=2.3]}
{$T/T_p$} & {$3p^5\,4s$} & {$\Delta$} & {$3p^5\,3d/5s$} & {$\Delta$} & {$3p^5\,4d$} & {$\Delta$} \\
\hline
\rule{0pt}{2.5ex}1 & 0.577 & {} & {} & {} & {} & {} \\
2 & 4.207 & 3.630 & 5.294 & {} & {} & {} \\
3 & 7.851 & 3.644 & 7.427 & 2.133 & {} & {} \\
4 & 11.600 & 3.749 & 8.611 & 1.184 & 3.899 & {} \\
5 & 15.201 & 3.600 & 12.386 & 3.774 & 6.062 & 2.163 \\
6 & 18.870 & 3.669 & 16.250 & 3.865 & 7.895 & 1.833 \rule[-0.2ex]{0pt}{0pt} \\ 
\end{tabular}
\end{ruledtabular}
\end{table}

Typical time-dependent calculations do not model decoherence (or spontaneous emission, which for most systems occurs on much longer timescales and is hence irrelevant) mechanisms, and thus these post-pulse oscillations continue indefinitely.
The corresponding spectral features increase and narrow systematically, with the linewidth scaling as \hbox{$\Delta\omega \sim 1/T$~\cite{Bracewell2000}}.
This is typically remedied by applying a window, but the ideal solution (at least as far as these effects are concerned) would be to ensure that the time window over which the Fourier transform of the dipole signal is taken matches that of the observation time in an associated experiment.

These results are quantified in Table~\ref{tab:int_scaled}, where the growth in the spectral fluence over time is shown for the first three ground-excited transitions.
Although the fourth and fifth features can be seen in the figure, they are not sufficiently developed (i.e., well-resolved) to properly quantify.
The lowest-energy peak corresponds to the $3p^5 4s\,{}^1P^o$ state in $LS$-coupling and can be matched closely to the NIST level at 11.828~eV with predominantly singlet character.
Since this peak is well isolated and has the largest spectral density, it demonstrates the expected linear increase in time most prominently.

The subsequent peak, centered around 14.1~eV in Fig.~\ref{fig:6}, cannot be uniquely assigned, as both the $3p^5 5s\,{}^1P^o$ and $3p^5 3d\,{}^1P^o$ states have experimental energies in this range (14.090~eV and 14.153~eV, respectively), within fine-structure splitting~\cite{kramida2024nist}.
It is therefore not surprising that the spectral fluence in this frequency range does not exhibit a clear linear growth at early times ($T < 4T_p$).
Furthermore, when linear growth does emerge at later times ($T \ge 4T_p$), it proceeds more rapidly than for the lower-energy (and larger) feature corresponding to the $3p^5 4s\,{}^1P^o$ state, owing to the overlap of the $3p^5 5s\,{}^1P^o$ and $3p^5 3d\,{}^1P^o$ contributions.
The subsequent two higher-energy features seen around 14.9 and 15.1~eV can be assigned to the $3p^5 4d\,{}^1P^o$ and $3p^5 6s\,{}^1P^o$ states.
These states require more time after the pulse to become sufficiently resolved in frequency before the expected linear growth can be quantified.

\section{Summary and Conclusions}

We have used the \textsc{RMT} method to carry out calculations for a rather extreme HHG regime in argon with a few-cycle, intense laser pulse, benchmarking against the joint experimental--theoretical study of Guo~\textit{et al.}~\cite{Guo2018}, who used very similar laser parameters.
Good agreement is found for the harmonic features reported above the ionization threshold, and the expected strong CEP dependence of the HHG spectra is also observed.

A central focus of the present study is the near-threshold excited-state region leading up to the first ionization threshold ($LS$-coupled) at 15.82~eV. 
In this region, care must be taken when presenting the HHG spectrum since the spectral density and fluence (total emitted radiation) depend substantially on the presence of windowing and post-pulse propagation time\,---\,both of which were investigated here.
Due to the population of bound states during the course of the pulse, coherent dipole oscillations ensue between the ground state (which has a 90\% survival rate in this particular case) and the accessible excited states.
This radiation manifests itself in spectral features that appear at the transition energies between these sets of states.
As the propagation extends beyond the pulse duration, we have confirmed in our simulation that these transition energies correspond to the field-free states and narrow with increasing observation time. 
However, applying the window to the dipole moment before taking its Fourier transform, a common practice in attosecond physics, suppresses these contributions that, at least in principle, are measurable.
This accumulation of signal occurs due to a continued emission of radiation after the laser pulse is turned off rather than a redistribution of a fixed spectral signal\,---\,windowing effectively rejects these late-time contributions.
We have demonstrated that both the Blackman and Tukey windows result in a disproportionately large change in spectral fluence for these near-threshold features.

Our work collectively demonstrates that the spectral density and fluence of harmonic features in the near-threshold region should be regarded\,---\,particularly on the theoretical side\,---\,as operational quantities that depend on the procedure used to obtain them.
Likewise, to compare \emph{quantitatively} with experiment, the effective observation time, determined by decoherence and detection, must be known.
The line strengths and linewidths of the features in this region are influenced by both the post-pulse propagation time and the presence of windowing. 
We recommend that, if the goal is to quantify the spectral fluence, windowing be avoided or used with great caution,  and that both theory and experiment quantify the temporal window over which the spectra are defined or measured.
The increase in the fluence of the near-threshold features reflects additional radiation from post-pulse coherent emission.
Experimentally, this time is often limited by decoherence processes, beyond which these persistent dipole oscillations are strongly suppressed.
In this case, a careful comparison between theory and experiment of the absolute and relative intensities of these features could be used as a means to extract the coherence time, which is experimentally influenced by a number of parameters such as collisions, focal averaging, and macroscopic effects.

In short-pulse regimes, such as analyzed in the present study, post-pulse dynamics have a major impact on the HHG response.
We have shown that a clear distinction can be made between these effects, which occur both during and after the pulse, and the more standard recollision-driven emission that is the hallmark of HHG, occurring only during the pulse.
Both processes are largely independent, though connected in that the bound-state population established by the pulse sets the initial condition for the subsequent coherent radiation.
In this picture, the quantities of most interest in attosecond physics, such as the plateau and cutoff, are still interpreted based on sub-cycle electron dynamics and the three-step model, while the near-threshold features are conceptually distinct, arising from field-free evolution of the wave function according to its population distribution.

In future work, simulations could incorporate decoherence and macroscopic propagation to limit the post-pulse coherence and thereby better mimic experimental observations.
Although we have shown that standard windowing techniques greatly suppress the near-threshold features, it would nevertheless be of interest to develop windowing procedures specifically designed, in concert with deliberate post-pulse propagation, to facilitate a quantitative analysis of these effects.
Achieving quantitative agreement between single-atom calculations and inherently macroscopic experiments is non\-trivial.  It will require proper accounting of both the microscopic coherence effects discussed here and the influence of macroscopic averaging.

\phantom{XYZ}

\section*{Acknowledgements}
We sincerely thank Prof.\ Anne Harth for clarifying the experimental aspects of the work reported in Ref.~\cite{Guo2018}, stimu\-lating discussions, and a careful reading of our initial manuscript with important suggestions for further improvement. We are also grateful to Prof.\ Kenneth Schafer for providing additional details regarding the calculations performed for the above paper.
This work was supported by the National Science Foundation under grants No.\  \hbox{PHY-2110023} and 
No.\ \hbox{PHY-2408484}, with computational resources provided by the ACCESS allocation PHY-090031 on Stampede-2 and Frontera at the Texas Austin Computing Center and on Expanse at the San Diego Super\-computing Center.

\bibliographystyle{apsrev4-2}
\bibliography{argon_hhg_rmt}

@article{clarke2018extreme,
  title={Extreme-ultraviolet-initiated high-order harmonic generation in Ar+},
  author={Clarke, D D A and van der Hart, H W and Brown, A C},
  journal={Phys. Rev. A},
  volume={97},
  number={2},
  pages={023413},
  year={2018},
  publisher={APS}
}

@article{hassouneh2018cooper,
  title={Cooper minimum in singly ionized and neutral argon},
  author={Hassouneh, O and Tyndall, N B and Wragg, J and van der Hart, H W and Brown, A C},
  journal={Phys. Rev. A},
  volume={98},
  number={4},
  pages={043419},
  year={2018},
  publisher={APS}
}

@book{Joachain2012,
  author = {Joachain, C. J. and Kylstra, N. J. and Potvliege, R. M.},
  title = {Atoms in Intense Laser Fields},
  publisher = {Cambridge University Press},
  year = {2012}
}

@article{telnov2013exterior,
  title={Exterior complex scaling method in time-dependent density-functional theory: Multiphoton ionization and high-order-harmonic generation of Ar atoms},
  author={Telnov, Dmitry A and Sosnova, Ksenia E and Rozenbaum, Efim and Chu, Shih-I},
  journal={Phys. Rev. A},
  volume={87},
  number={5},
  pages={053406},
  year={2013},
  publisher={APS}
}

@book{Jackson1999,
  author = {Jackson, J. D.},
  title = {Classical Electrodynamics},
  edition = {3rd},
  publisher = {Wiley},
  year = {1999}
}

@article{Han2010,
  author = {Y.-C. Han and L. B. Madsen},
  title = {Agreement of gauge-dependent forms in strong-field calculations},
  journal = {Phys. Rev. A},
  volume = {81},
  pages = {063430},
  year = {2010}
}

@article{Bandrauk2009,
  title = {Quantum simulation of high-order harmonic spectra of the hydrogen atom},
  author = {Bandrauk, A. D. and Chelkowski, S. and Diestler, D. J. and Manz, J. and Yuan, K.-J.},
  journal = {Phys. Rev. A},
  volume = {79},
  issue = {2},
  pages = {023403},
  numpages = {14},
  year = {2009},
  month = {Feb},
  publisher = {American Physical Society},
  doi = {10.1103/PhysRevA.79.023403},
  url = {https://link.aps.org/doi/10.1103/PhysRevA.79.023403}
}

@article{Brown2020,
  title={RMT: R-matrix with time-dependence. Solving the semi-relativistic, time-dependent Schr{\"o}dinger equation for general, multielectron atoms and molecules in intense, ultrashort, arbitrarily polarized laser pulses},
  author={Brown, Andrew C and Armstrong, Gregory SJ and Benda, Jakub and Clarke, Daniel DA and Wragg, Jack and Hamilton, Kathryn R and Ma{\v{s}}{\'\i}n, Zden{\v{e}}k and Gorfinkiel, Jimena D and van der Hart, Hugo W},
  journal={Comp. Phys. Commun.},
  volume={250},
  pages={107062},
  year={2020},
  publisher={Elsevier}
}

@article{Corkum1993,
  author = {Corkum, P. B.},
  title = {Plasma perspective on strong-field multiphoton ionization},
  journal = {Phys. Rev. Lett.},
  volume = {71},
  pages = {1994--1997},
  year = {1993}
}

@article{Lewenstein1994,
  author = {Lewenstein, M. and Balcou, P. and Ivanov, M. Y. and L'Huillier, A. and Corkum, P. B.},
  title = {Theory of high-harmonic generation by low-frequency laser fields},
  journal = {Phys. Rev. A},
  volume = {49},
  pages = {2117--2132},
  year = {1994}
}

@article{Paul2001,
  author = {Paul, P. M. and Toma, E. S. and Breger, P. and Mullot, G. and Aug{\'e}, F. and Balcou, P. and Muller, H. G. and Agostini, P.},
  title = {Observation of a train of attosecond pulses from high harmonic generation},
  journal = {Science},
  volume = {292},
  pages = {1689--1692},
  year = {2001}
}

@article{Bondy2024high,
  title={High-order harmonic generation in helium: A comparison study},
  author={Bondy, A. T. and Saha, S and Del Valle, J. C. and Harth, Anne and Douguet, N and Hamilton, K R and Bartschat, K},
  journal={Phys. Rev. A},
  volume={109},
  number={4},
  pages={043113},
  year={2024},
  publisher={APS}
}

@article{Pauly2020,
  author = {Pauly, T. and Bondy, A. T. and Hamilton, K. R. and Douguet, N. and Tong, X.-M. and Chetty, D. and Bartschat, K.},
  title = {Ellipticity dependence of excitation and ionization of argon atoms by short-pulse infrared radiation},
  journal = {Phys. Rev. A},
  volume = {102},
  pages = {013116},
  year = {2020}
}

@incollection{Schafer2009,
  author    = {Schafer, Kenneth J.},
  title     = {Strong-Field Physics},
  booktitle = {Strong Field Laser Physics},
  editor    = {Brabec, Thomas},
  publisher = {Springer},
  year      = {2009},
  pages     = {111--145},
  doi       = {10.1007/978-0-387-34755-4_6}
}

@article{KrauszIvanov2009,
  author = {Krausz, F. and Ivanov, M.},
  title = {Attosecond physics},
  journal = {Rev. Mod. Phys.},
  volume = {81},
  pages = {163--234},
  year = {2009}
}

@article{Higuet2011,
  author = {Higuet, J. and Ruf, H. and Thir{\'e}, N. and Cireasa, R. and Constant, E. and Cormier, E. and Descamps, D. and M{\'e}vel, E. and Petit, S. and Pons, B. and Mairesse, Y. and Fabre, B.},
  title = {High-order harmonic spectroscopy of the Cooper minimum in argon: Experimental and theoretical study},
  journal = {Phys. Rev. A},
  volume = {83},
  pages = {053401},
  year = {2011}
}

@article{Guo2018,
  title={Phase control of attosecond pulses in a train},
  author={Guo, Chen and Harth, Anne and Carlstr{\"o}m, Stefanos and Cheng, Yu-Chen and Mikaelsson, Sara and M{\aa}rsell, Erik and Heyl, Christoph and Miranda, Miguel and Gisselbrecht, Mathieu and Gaarde, Mette B and others},
  journal={J. Phys. B: At. Mol. Opt. Phys.},
  volume={51},
  number={3},
  pages={034006},
  year={2018},
  publisher={IOP Publishing}
}

@article{burke1975r,
  title={R-matrix theory of photoionization. Application to neon and argon},
  author={Burke, P. G. and Taylor, K. T.},
  journal={J. Phys. B: At. Mol. Phys.},
  volume={8},
  number={16},
  pages={2620--2639},
  year={1975}
}

@article{Finger2022,
  author = {Finger, K. and Atri-Schuller, D. and Douguet, N. and Bartschat, K. and Hamilton, K. R.},
  title = {High-order harmonic generation in the water window from mid-{IR} laser sources},
  journal = {Phys. Rev. A},
  volume = {106},
  pages = {063113},
  year = {2022}
}

@article{Camp2015,
  author = {Camp, S. and Schafer, K. J. and Gaarde, M. B.},
  title = {Interplay between resonant enhancement and quantum path dynamics in harmonic generation in helium},
  journal = {Phys. Rev. A},
  volume = {92},
  pages = {013404},
  year = {2015}
}

@article{Beaulieu2016,
  author = {Beaulieu, S. and Comtois, D. and Bond, E. and Boudreau, S. and Desc{\^o}teaux, S. and Khalili, K. and Schouder, C. and Azarm, A. and Schmidt, B. E. and L{\'e}gar{\'e}, F. and Villeneuve, D. M. and W{\"o}rner, H. J.},
  title = {Role of excited states in high-order harmonic generation},
  journal = {Phys. Rev. Lett.},
  volume = {117},
  pages = {203001},
  year = {2016}
}

@book{Bracewell2000,
  author = {Bracewell, R. N.},
  title = {The Fourier Transform and Its Applications},
  publisher = {McGraw-Hill},
  year = {2000}
}

@article{Yun2018,
  author = {Yun, H. and Mun, J. H. and Hwang, S. I. and Park, S. B. and Ivanov, I. A. and Nam, C. H. and Kim, K. T.},
  title = {Coherent extreme-ultraviolet emission generated through frustrated tunnelling ionization},
  journal = {Nat. Photon.},
  volume = {12},
  pages = {620--624},
  year = {2018}
}

@book{BlackmanTukey1959,
  author    = {R. B. Blackman and J. W. Tukey},
  title     = {The Measurement of Power Spectra: From the Point of View of Communications Engineering},
  publisher = {Dover},
  address   = {New York},
  year      = {1959},
  pages     = {98--99}
}

@article{Harris1978,
  author  = {F. J. Harris},
  title   = {On the use of windows for harmonic analysis with the discrete Fourier transform},
  journal = {Proc. IEEE},
  volume  = {66},
  number  = {1},
  pages   = {51--83},
  year    = {1978},
  doi     = {10.1109/PROC.1978.10837}
}

@article{Kulander1991,
  author = {Kulander, K. C. and Rescigno, T. N.},
  title = {Effective potentials for time-dependent calculations of multiphoton processes in atoms},
  journal = {Comput. Phys. Commun.},
  volume = {63},
  pages = {523--528},
  year = {1991}
}

@article{Tong2005,
  author = {Tong, X. M. and Lin, C. D.},
  title = {Empirical formula for static field ionization rates of atoms and molecules by lasers in the barrier-suppression regime},
  journal = {J. Phys. B: At. Mol. Opt. Phys.},
  volume = {38},
  pages = {2593--2600},
  year = {2005}
}

@article{Mauritsson2005,
  author = {Mauritsson, J. and Gaarde, M. B. and Schafer, K. J.},
  title = {Accessing properties of electron wave packets generated by attosecond pulse trains through time-dependent calculations},
  journal = {Phys. Rev. A},
  volume = {72},
  pages = {013401},
  year = {2005}
}

@misc{kramida2024nist,
  title={{NIST} {A}tomic {S}pectra {D}atabase (ver.~5.12)},
  author={Kramida, A and Ralchenko, Yu and Reader, J and others},
  year={2024},
  publisher={National Institute of Standards and Technology Gaithersburg, MD}
}

@article{peng2025resolving,
  title={Resolving Recapture Dynamics of Rydberg Electrons via Laser-Driven Frustrated Tunneling Ionization},
  author={Peng, Sainan and Chen, Yudong and Li, Yang and Fan, Guangyu and Xie, Xinhua and He, Feng and Tao, Zhensheng},
  journal={Phys. Rev. Lett.},
  volume={134},
  number={12},
  pages={123203},
  year={2025},
  publisher={APS}
}

@article{Gibson2004,
  author = {Gibson, E. A. and Paul, A. and Wagner, N. and Tobey, R. and Gaudiosi, D. and Backus, S. and Christov, I. P. and Kapteyn, H. C. and Murnane, M. M.},
  title = {High-order harmonic generation up to 250 eV from highly ionized argon},
  journal = {Phys. Rev. Lett.},
  volume = {92},
  pages = {033001},
  year = {2004}
}

\end{document}